\documentclass{llncs}
\usepackage[T1]{fontenc}
\usepackage{tgtermes}
\usepackage{url}

\usepackage[normalem]{ulem}

\usepackage[linesnumbered,ruled,vlined]{algorithm2e}
\usepackage{graphicx}

\usepackage{booktabs}
\usepackage{multirow}

\author{
Yuliang Li$^1$ \and
Alin Deutsch$^1$ \and
Victor Vianu$^1$ $^2$
} 

\institute{
\begin{tabular}{cc}
$^1$University of California, San Diego &
\hspace*{1mm} $^2$INRIA Paris  \\
  {\normalsize\tt \{yul206,deutsch,vianu\}@cs.ucsd.edu} \hspace*{1mm} &   
\end{tabular}
}

\authorrunning{Y. Li et. al.} 

\title{SpinArt: A Spin-based Verifier for Artifact Systems}

\usepackage{graphicx}

\newcommand{\type}{\ensuremath{\mathtt{type}}}

\usepackage{color}

\newcommand{\alin}[1]{{\it\small\textcolor{green}{[[[ {#1}\ --alin ]]]}}}
\newcommand{\victor}[1]{{\it\small\textcolor{red}{[[[ {#1}\ --victor ]]]}}}
\newcommand{\yuliang}[1]{{\it\small\textcolor{blue}{[[[ {#1}\ --yuliang ]]]}}}
\newcommand{\reviewer}[1]{{\it\small\textcolor{cyan}{[[[ {#1}\ --reviewers ]]]}}}
\renewcommand{\alin}[1]{}
\renewcommand{\victor}[1]{}
\renewcommand{\yuliang}[1]{}
\renewcommand{\reviewer}[1]{}

\newcommand{\eat}[1]{}

\newcommand{\cala}{{\cal A}}

\newcommand{\calp}{{\cal P}}

\newcommand{\cale}{{\cal E}}

\newcommand{\lora}{\longrightarrow}
\newcommand{\goto}[1]{\stackrel{#1}{\lora}}
\newcommand{\db}{\mathcal{DB}}

\newcommand{\anull}{\ensuremath{\mathtt{null}}}

\newcommand{\varid}{\ensuremath{\emph{VAR}_{id}}}
\newcommand{\varnum}{\ensuremath{\emph{VAR}_{\emph{val}}}}

\newcommand{\dbcustomers}{\ensuremath{\mathtt{CUSTOMERS}}}
\newcommand{\dbitems}{\ensuremath{\mathtt{ITEMS}}}
\newcommand{\dbrecords}{\ensuremath{\mathtt{CREDIT\_RECORD}}}

\newcommand{\acustomerid}{\ensuremath{\mathtt{cust\_id}}}
\newcommand{\aitemid}{\ensuremath{\mathtt{item\_id}}}
\newcommand{\astatus}{\ensuremath{\mathtt{status}}}

\newcommand{\ainstock}{\ensuremath{\mathtt{instock}}}

\makeatletter
\def\footnoterule{\kern-3\p@
  \hrule \@width 2in \kern 2.6\p@} 
\makeatother

\usepackage{listings}
\usepackage{lipsum}
\usepackage{courier}
\usepackage{amsmath}

\begin{document}
\maketitle

\begin{abstract}
Data-driven workflows, of which IBM's Business Artifacts are a prime exponent, 
have been successfully deployed in practice, adopted in industrial standards, and
have spawned a rich body of research in academia, focused primarily on static analysis. 
In previous work, theoretical results were obtained on the verification of a rich model 
incorporating core elements of IBM's successful Guard-Stage-Milestone (GSM) artifact model. 
The results showed decidability of verification of temporal properties of a large 
class of GSM workflows and established its complexity. Following up on these results, 
the present paper reports on the implementation of SpinArt,
a practical verifier based on the classical model-checking tool Spin.
The implementation includes nontrivial optimizations and achieves good performance on real-world business 
process examples. Our results shed light on the capabilities and limitations of off-the-shelf verifiers in the 
context of data-driven workflows.
\end{abstract}


\section{Introduction} \label{sec:intro}
The past decade has witnessed the evolution of workflow specification frameworks from the traditional
process-centric approach towards data-awareness. Process-centric formalisms focus on control flow 
while under-specifying the underlying data and its manipulations by the process
tasks, often abstracting them away completely. In contrast, data-aware formalisms treat data as first-class
citizens. A notable exponent of this class is IBM's {\em business artifact model} 
pioneered in~\cite{Nigam03:artifacts}, successfully deployed in practice ~\cite{Bhatt-2005:Artifacts-pharm,Bhatt-2007:artifacts-customer-engagements,IGF-case-study:BPM-2009,Cordys-case-management,IBM-case-mgmt} and adopted in industrial standards.

In a nutshell, business artifacts (or simply ``artifacts'') model key business-relevant
entities, which are updated by a set of services that implement
business process tasks, specified declaratively by pre-and-post conditions. 
A collection of artifacts and services is called an {\em artifact system}. 
IBM has developed several variants of artifacts, of which the most recent is 
Guard-Stage-Milestone (GSM) \cite{Damaggio:BPM11,GSM:DEBS-2011}. 
The GSM approach provides rich structuring mechanisms for services, including 
parallelism, concurrency and hierarchy, and has been incorporated in the OMG standard 
for Case Management Model and Notation (CMMN) \cite{OMG:CMMN:Beta1,GSM-CMMN:2012}.

Artifact systems deployed in industrial settings typically specify complex workflows 
prone to costly bugs, whence the need for verification of critical properties.
Over the past few years, the verification problem for artifact systems was intensively studied.
The focus of the research community has been to identify practically relevant classes of artifact systems and 
properties for which {\em fully automatic} verification is possible.
This is an ambitious goal, since artifacts are infinite-state systems due to the presence of unbounded data. 
Along this line, complexity results were shown for different versions of 
the verification problem with various expressiveness of the artifact models and properties.
In particular, a previous work \cite{pods16} studied Hierarchical Artifact Systems (HAS), 
a model capturing core elements of the GSM model, and established the complexity of 
verifying a rich class of linear-time temporal properties for various fragments of HAS.

The present paper follows up on the theoretical results of \cite{pods16} by 
studying the practical implementation of \emph{SpinArt}, a fully automatic verifier for artifact systems.
The goal in this work is to explore the feasibility of using existing off-the-shelf tools
to implement such an artifact verifier. We focus specifically on Spin \cite{spin}, the main 
model checker used in the verification community and the natural candidate for a verifier implementation.

We begin by defining a core fragment of the HAS model, called \emph{Tuple Artifact System} (TAS), 
that can potentially be handled by Spin.
At a high level, a tuple artifact system consists of a read-only database, a tuple of updatable artifact variables
and a set of \emph{services} specifying transitions of the system using pre-and-post conditions. 
This fragment remains very expressive, as demonstrated by our experiments showing that
a large set of realistic business processes can be specified as TAS's.
The properties of TAS's to be verified are specified using an extension of Linear-Time Temporal Logic (LTL).

Our model is expressive enough to allow data of unbounded domain and size, 
which are features not directly supported by Spin or other state-of-the-art model checkers. 
Therefore, a direct translation into Spin requires setting limits on the size of the data and its domain,
resulting in an incomplete verifier.
To address this challenge, we exploit the symbolic verification techniques establishing
the decidability results in \cite{pods16} and develop a simple algorithm for translating
TAS specifications and properties into equivalent problem instances that can be verified by Spin, 
{\em without sacrificing either the soundness or the completeness of the verifier}.
However, a naive use of Spin still results in poor performance even with the translation algorithm. Therefore, we develop
an array of nontrivial optimizations techniques to render verification tractable.
To the best of our knowledge, SpinArt is the first implementation of an 
artifact system verifier that preserves decidability under unbounded data while being
based on off-the-shelf model checking technology.
The main contributions are summarized as follows.
\begin{itemize}\parskip=0in\itemsep=0in
\item 
We define Tuple Artifact System (TAS), 
a core fragment of HAS that permits efficient implementation of a Spin-based verifier.
By exploiting the symbolic verification approach from previous work \cite{pods16,tods12}, 
we show a simple algorithm for translating the verification problem into an equivalent instance in Spin.
This algorithm forms the basis of our implementation of SpinArt.
\item 
We implement SpinArt with two nontrivial optimization techniques
to achieve satisfactory performance.
The first consists of a more efficient translation algorithm avoiding a quadratic blowup
in the size of the specification due to keys and foreign keys, 
so that it shortens significantly the compilation and execution time for Spin.
The second optimization is based on static analysis, and greatly reduces
the size of the search space by exploiting constraints extracted from the input specification 
during a pre-computation phase.
Although these techniques are designed with Spin as the target tool,
we believe that they can be adapted to implementations based on other off-the-shelf model checkers.
\item
We evaluate the performance of SpinArt experimentally using both real-world and synthetic data-driven workflows
and properties. 
We created a benchmark of artifact systems and LTL-FO properties 
from existing sets of business process specifications and temporal properties by extending them 
with data-aware features. The experiments highlight the impact of the optimizations and various
parameters of the specifications and properties on the performance of SpinArt. 
\end{itemize}

The paper is organized as follows.
We start by reviewing in Sect. \ref{sec:newmodel} the HAS model and 
formally defining TAS, a core fragment of HAS. 
We also review LTL-FO, the temporal logic for specifying properties of TAS's. 
In Sect. \ref{sec:spin} we first review the theory developed in \cite{pods16}, then 
describe the initial direct implementation of SpinArt 
based on the symbolic representation technique introduced there. 
We next present the specialized optimizations, essential for achieving acceptable performance. 
The experimental results are shown in Sect. \ref{sec:experiment}. 
Finally, we discuss related work in Sect. \ref{sec:related} and conclude in Sect. \ref{sec:conclusion}.

\section{The Model}\label{sec:newmodel}

In this section, we present the variant of artifact systems supported by our verifier, as well as
the temporal logic LTL-FO used to specify the properties to be verified.

\subsection{Tuple Artifact Systems}

The model is a variant of the Hierarchical Artifact System (HAS) model
presented in \cite{pods16}.  
In brief, a HAS consists of a database and a hierarchy (rooted tree) of {\em tasks}.
Each task has associated to it local evolving data consisting of a tuple of artifact variables and an updatable 
set of tuples called the artifact relation.
It also has an associated set of {\em services}.
Each application of a service is guarded by a pre-condition on the database and local data and
causes an update of the local data, specified by a post condition (constraining the next artifact tuple)
and an insertion or retrieval of a tuple from the artifact relation.
In addition, a task may invoke a child task with a tuple of parameters,
and receive back a result when the child task completes. A run of the artifact system is
obtained by any valid interleaving of concurrently running task services.

The implemented model restricts the HAS model as follows:
\begin{itemize}\itemsep=0pt\parskip=0pt
\item it disallows evolving relations in artifact data
\item it does not use arithmetic in service pre-and-post conditions
\item the underlying database schema uses an {\em acyclic} set of foreign keys\footnote{Foreign keys and acyclic schemas 
are standard database notions, reviewed in Definition \ref{def:schema}.} 
\end{itemize}
As shown by the real-life examples used in the experimental evaluation, 
the implemented model is powerful enough to capture a wide variety of business processes,
and is a good vehicle for studying the implementation of a Spin-based verifier. 

The implemented model retains the hierarchy of tasks present in HAS. However, for simplicity of exposition, we 
only define formally the core of the model, consisting of a single task in which
a tuple of artifact values evolves throughout the workflow under the action of services. 
For clarity, we also describe the algorithms in terms of the core model.
The exposition can be easily extended to a hierarchy of tasks.

   
We now present the syntax and semantics of the core model, which we call \emph{Tuple Artifact System} (TAS). 
The formal definitions below are illustrated with an intuitive example of the TAS specification of
an order fulfillment business process originally written in BPMN \cite{bpmn}.
Intuitively, the workflow allows customers to place orders and the supplier company to process the orders.

We begin with the underlying database schema. 
\vspace*{-1mm}
\begin{definition} \label{def:schema}
A \textbf{database schema} $\db$ is a finite set of relation symbols, where 
each relation $R$ of $\db$ has an associated sequence of distinct attributes 
containing the following: 

\vspace{-3mm}
\begin{itemize}\itemsep=0pt\parskip=0pt
\item a key attribute $\emph{ID}$ (providing a unique identifier for tuples in $R$), 
\item a set of foreign key attributes $\{F_1, \dots, F_m\}$, and
\item a set of non-key attributes $\{A_1, \dots, A_n\}$ disjoint from $\{\emph{ID}, F_1, \dots, F_m\}$.
\end{itemize}
\vspace{-1mm}
To each foreign key attribute $F_i$ of $R$ is associated a relation $R_{F_i}$ of $\db$
and the inclusion dependency $R[F_i] \subseteq R_{F_i}[\emph{ID}]$, stating that every value of attribute $F_i$ occurring in $R$
is the ID of a tuple in $R_{F_i}$.  It is said that the foreign key $F_i$ references relation $R_{F_i}$.
\end{definition} 
\vspace*{-1mm}

Intuitively, a foreign key $F$ of relation $R$ referencing relation $R_F$ acts as a pointer from the tuples of $R$ to tuples of $R_F$.
The assumption that the ID of each relation is a single attribute is made for simplicity, and multiple-attribute IDs
can be easily handled.  

A database schema $\db$ is {\em acyclic} if there are no cycles in the references induced by foreign keys.
More precisely, consider the directed graph FK whose nodes are the relations of the schema and
in which there is an edge from $R_i$ to $R_j$ if
$R_i$ has a foreign key attribute $F$ referencing $R_j$.
The schema $\db$ is {\em acyclic} if the graph FK is acyclic.
All database schemas considered in this paper are acyclic.
Note that acyclic schemas include the Star (and Snowflake) schemas \cite{starschema1,starschema2} widely used in business process data management.

\vspace*{-1mm}
\begin{example} \label{exm:db}
The order fulfillment workflow has the following database schema:
\vspace*{-1mm}
\begin{itemize}\itemsep=0pt\parskip=0pt
\item
\dbcustomers $\mathtt{({\emph{ID}}, name, address, record)}$, 
$\quad$\dbitems $\mathtt{({\emph{ID}}, item\_name, price)}$\\
\dbrecords $\mathtt{({\emph{ID}}, status)}$\\
\end{itemize}
\vspace*{-5mm}

\noindent
The \emph{ID}s are key attributes, 
$\mathtt{price}$, $\mathtt{item\_name}$, $\mathtt{name}$, $\mathtt{address}$, $\mathtt{status}$
are non-key attributes, and $\mathtt{record}$ is a foreign key attribute
satisfying the dependency $\dbcustomers[record]$ $\subseteq \dbrecords[\emph{ID}]$.
Intuitively, the $\dbcustomers$ table contains customer information with 
a foreign key pointing to the customers' credit records stored in $\dbrecords$. 
The $\dbitems$ table contains information on the items. 
Note that the schema is acyclic as there is only one foreign key reference from $\dbcustomers$ to $\dbrecords$.
\end{example} 
\vspace*{-1mm}

We assume two infinite, disjoint domains of IDs and data values, 
denoted by $\emph{DOM}_{id}$ and $\emph{DOM}_{val}$, and an additional
constant $\anull$ where $\anull \not\in \emph{DOM}_{id} \cup \emph{DOM}_{val}$
($\anull$ is useful as a special initialization value).
The domain of all non-key attributes is  $\emph{DOM}_{val}$. 
The domain of each key attribute ID of relation $R$ is an infinite subset 
$Dom(R.\emph{ID})$ of $\emph{DOM}_{id}$, and
$Dom(R.\emph{ID}) \cap Dom(R'.\emph{ID}) = \emptyset$ for $R \neq R'$. 
The domain of a foreign key attribute $F$ referencing $R$ is $Dom(R.\emph{ID})$. 
Intuitively, in such a database schema, each
tuple is an object with a \emph{globally} unique id. This id does not appear
anywhere else in the database except as foreign keys referencing it.
An {\em instance} of a database schema $\db$ is a mapping $D$ associating to each relation symbol $R$ a finite relation
$D(R)$ of the same arity of $R$, whose tuples provide, for each attribute, a value from its domain, such that no distinct tuples
agree on the key \emph{ID}. In addition,
$D$ satisfies all inclusion dependencies associated with the foreign keys of the schema. 
\yuliang{removed the active domain since it is not used in this paper.}

\vspace*{-2mm}
\begin{example} \label{exm:db-instance}
Figure \ref{fig:instance} shows an example of an instance of 
the acyclic schema of the order fulfillment workflow.
Note that the domains of $\dbcustomers$.\emph{ID}, $\dbitems$.\emph{ID} and $\dbrecords$.\emph{ID}
and the domain for non-key attributes are mutually disjoint.
The domain of $\dbcustomers$.record is included in $Dom(\dbrecords.\emph{ID})$ since
$\mathtt{record}$ is a foreign key attribute referencing $\dbrecords.\emph{ID}$.
\begin{figure}[!ht]
\vspace*{-6mm}
\centering
\includegraphics[width=0.9\textwidth]{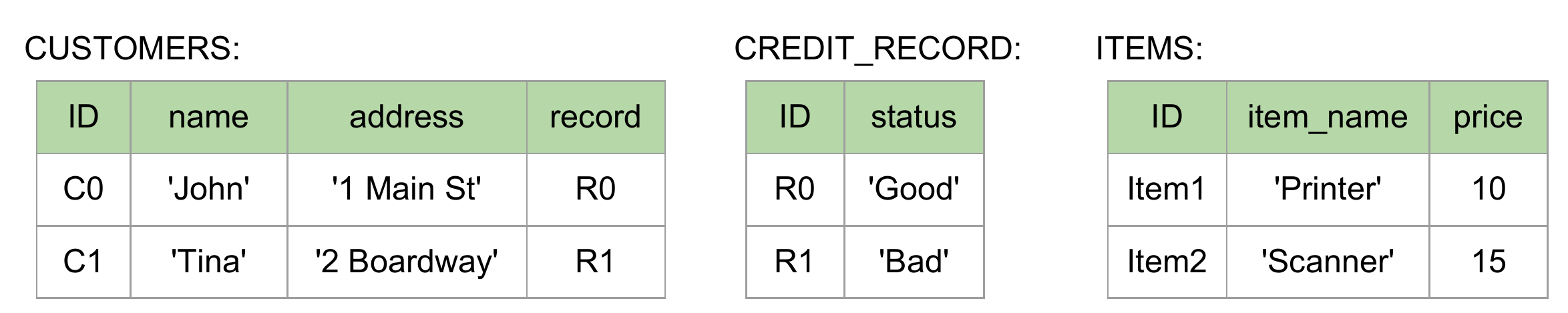}
\caption{An instance of an acyclic schema.}
\label{fig:instance}
\end{figure}
\end{example}
\vspace*{-2mm}

\noindent
We next proceed with the definition of artifacts and services.
Similarly to the database schema, we consider two infinite, disjoint sets $\varid$ of ID variables and $\varnum$ of
data variables. We associate to each variable $x$ its domain $Dom(x)$.
If $x \in \varid$, then $Dom(x) = \emph{DOM}_{id} \cup \{\anull\}$, 
and if $x \in \varnum$, then $Dom(x) = \emph{DOM}_{val} \cup \{\anull\}$. 
An {\em artifact variable} is a variable in $\varid \cup \varnum$.
If $\bar{x}$ is a sequence of artifact variables, a {\em valuation} of $\bar{x}$ 
is a mapping $\nu$ associating to each variable $x$ in $\bar{x}$ an element in $Dom(x)$. 

\vspace{-1mm}
\begin{definition}
An \textbf{artifact schema} is a pair $\cala = \langle \db, \bar{x} \rangle$ with an acyclic database schema $\db$
and $\bar{x} \subseteq \varid \cup \varnum$ a set of artifact variables.
The domain of each variable $x \in \bar{x}$ is either 
$\emph{DOM}_{val} \cup \{\anull\}$ or $dom(R.\emph{ID}) \cup \{\anull\}$ for some relation $R \in \db$.
In the latter case we say that the type of $x$ is  $\type(x) = R.\emph{ID}$.
An \textbf{instance} $\rho$ of $\cala$ is a pair $(D, \nu)$ where 
$D$ is a finite instance of $\db$ and $\nu$ is a valuation of $\bar{x}$.
\end{definition}
\vspace{-1mm}

\vspace{-1mm}
\begin{example} \label{exm:schema}
The artifact schema of the order fulfillment example consists of 
the acyclic database schema described in Example \ref{exm:db} and the following artifact variables:
\vspace{-1mm}
\begin{itemize}\itemsep=0pt\parskip=0pt
\item ID variables: $\acustomerid$ of type $\dbcustomers.\emph{ID}$ and $\aitemid$ of type $\dbitems.\emph{ID}$
\item Non-ID variables: $\astatus$ and $\ainstock$
\end{itemize}
\vspace{-1mm}
Intuitively, $\acustomerid$ and $\aitemid$ store the ID of the customer and 
the ID of the item ordered by the customer. Variable $\astatus$ indicates 
the different stages of the order, namely ``Init'', ``OrderPlaced'', ``Passed'' (passed the credit check),
``Shipped'' or ``Failed''. Variable $\ainstock$ indicates whether the ordered item is in stock.
\end{example}
\vspace{-2mm}

For a given artifact schema $\cala = \langle \db, \bar{x} \rangle$
and a sequence $\bar y$ of variables, a {\em condition} on $\bar y$ is 
a quantifier-free first-order (FO) formula over 
$\mathcal{DB} \cup \{=\}$ whose variables are included in $\bar{y}$. In more detail, a condition over $\bar y$ is 
a Boolean combination of relational or equality atoms whose variables are included in $\bar y$.  
A relational atom over relation 
$R(\emph{ID}, A_1, \dots, A_m, F_1, \dots, F_n) \in \db$, 
is of the form  $R(x, y_1, \dots, y_m, z_1, \dots, z_n)$, 
where $\{x, z_1, \dots, z_n\} \subseteq \varid$ and $\{y_1, \dots, y_m\} \subseteq \varnum$.
An equality atom is of the form $x = z$, where $x$ is variable and $z$ is a variable of the same type, or $x \in \varnum$
and $z \in \emph{DOM}_{val}$. 
The special constant $\anull$ can be used in equalities.
If $\alpha$ is a condition on $\bar y \subseteq \bar x$, 
$D$ an instance of $\mathcal{DB}$ and $\nu$ a valuation of $\bar x$, 
we denote by $D \models \alpha(\nu)$ the fact that $D$ satisfies $\alpha$ with valuation $\nu$, with standard semantics. 
For an atom $R(\bar{z})$ in $\alpha$ where $R \in \db$, 
if $\nu(z) = \anull$ for some $z \in \bar{z}$, then $R(\nu(\bar{z}))$ is false (since the database 
instances do not contain $\anull$).
Although conditions are quantifier-free, conditions with existentially quantified variables (denoted $\exists$FO) 
can be easily simulated by adding variables to $\bar{x}$, so we use them as shorthand whenever convenient.

\vspace{-1mm}
\begin{example}\label{exm:condition}
The following $\exists$FO condition 
states that the customer with ID $\acustomerid$ has good credit:
\vspace*{-2mm}
$$\quad \exists n \exists a \exists r \ 
\dbcustomers(\acustomerid, n, a, r) \land \dbrecords(r, \text{``Good''}).$$
\end{example}
\vspace{-1mm}
We next define services in TAS. 
\vspace{-1mm}
\begin{definition}
Let $\cala = \langle \db, \bar{x} \rangle$ be an artifact schema.
A \textbf{service} $\sigma$ of $\cala$ is a tuple $\langle \pi, \psi, \bar y \rangle$ where:
\vspace*{-1mm}
\begin{itemize}\itemsep=0pt\parskip=0pt
\item $\pi$ and $\psi$, called \emph{pre-condition} and \emph{post-condition}, 
respectively, are conditions over $\bar{x}$, and
\item $\bar y$ is the set of {\em propagated variables}, where $\bar{y} \subseteq  \bar{x}$.
\end{itemize}
\end{definition}
\vspace{-1mm}
Intuitively, $\pi$ and $\psi$ are conditions which must be satisfied by the previous and the next instance respectively when $\sigma$ is applied. 
In addition, the values stored in $\bar y$ are propagated to the next instance. 

\vspace{-1mm}
\begin{example}\label{exm:service}
The order fulfillment TAS has the following five services: 
\textbf{EnterCustomer}, \textbf{EnterItem}, \textbf{CheckCredit}, \textbf{Restock} and \textbf{ShipItem}.
Intuitively, for each order, the workflow first obtains the customer and 
item information by applying the \textbf{EnterCustomer} service and the \textbf{EnterItem} service. 
Then the credit record of the customer is checked by the \textbf{CheckCredit} service. 
If the record is good, \textbf{ShipItem} can be called to ship the item to the customer. 
If the requested item is unavailable,
then \textbf{Restock} must be called before \textbf{ShipItem} to procure the item.

Next, we illustrate each service in more detail. 
The \textbf{EnterCustomer} and \textbf{EnterItem} allow the customer to enter 
his/her information and the ordered item's information. The $\dbcustomers$ and $\dbitems$
tables are queried to obtain the customer ID and item ID. 
When \textbf{EnterItem} is called, the supplier also checks whether the item is currently in stock and
sets the variable $\ainstock$ to ``Yes'' or ``No'' accordingly.
This step is modeled as an external service so we use the post-condition to
enforce that the two values are chosen nondeterministically.
In both services, if both $\acustomerid$ and $\aitemid$ have been entered,
the current status of the order is updated to ``OrderPlaced'' (otherwise it remains ``Init'').
The two services can be called multiple times to 
allow the customer to modify previously entered data. 
The propagated variables of \textbf{EnterCustomer} are $\aitemid$ and $\ainstock$ since their values
are not modified when the service is applied. 
Similarly, the only propagated variable of \textbf{EnterItem} is $\acustomerid$.
The two services are formally specified in Fig. \ref{fig:services},
and Fig. \ref{fig:transition} shows transitions that result from applying the two services consecutively.
\yuliang{Need to make sure the figure are in the right place}

\begin{figure}[!ht]
\noindent\begin{minipage}{.5\textwidth}
\noindent
\textbf{EnterCustomer}:  \\
Pre-condition: $\astatus = \text{``Init''}$ \\ 
Propagated: $\{\aitemid, \ainstock\}$ \\
Post-condition:
\begingroup
\addtolength{\jot}{-0.7mm}
\begin{align*}
& \exists n \exists a \exists r \ \dbcustomers(\acustomerid, n, a, r) \land \quad\quad \\
& (\aitemid \neq \anull \rightarrow \\
& \astatus = \text{``OrderPlaced"} ) \land \\
& (\aitemid = \anull \rightarrow \astatus = \text{``Init''} ) 
\end{align*}
\endgroup
\end{minipage}\hfill
\begin{minipage}{.5\textwidth}
\noindent
\textbf{EnterItem}:  \\
Pre-condition: $\astatus = \text{``Init''}$ \\ 
Propagated: $\{\acustomerid\}$ \\
Post-condition:
\begingroup
\addtolength{\jot}{-0.7mm}
\begin{align*}
&  \exists n \exists p \ \dbitems(\aitemid, n, p) \land \\
& (\ainstock = \text{``Yes'' } \lor \ainstock = \text{``No''}) \land \\
& (\acustomerid \neq \anull \rightarrow \astatus = \text{``OrderPlaced"} ) \land \\
& (\acustomerid = \anull \rightarrow \astatus = \text{``Init''} )
\end{align*}
\endgroup
\end{minipage}
\caption{Examples of two services.}\label{fig:services}
\end{figure}



\begin{figure}[!ht]
\vspace{5mm}
\centering
\includegraphics[width=1\textwidth]{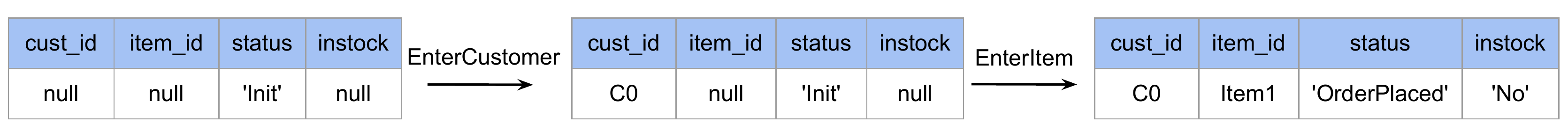}
\caption{Two transitions caused by services.}
\label{fig:transition}
\end{figure}

We describe in brief the rest of the services.
The \textbf{CheckCredit} service can be called if $\astatus = \text{``OrderPlaced''}$.
It checks the credit record of the customer using the condition $\text{IsGood(\acustomerid)}$ in Example \ref{exm:condition}. If the credit record is good, then it updates $\astatus$ to ``Passed'' otherwise to ``Failed''.
The \textbf{Restock} service can be called if $\astatus = \text{``Passed''}$ which means that
the credit check is passed. The service simply updates $\ainstock$ to ``Yes'', 
indicating that ordered item is now in stock.
Finally, the \textbf{ShipItem} can be called if $\astatus = \text{``Passed''}$ and $\ainstock = \text{``Yes''}$.
It updates $\astatus$ to ``Shipped'', meaning that the shipment is successful.
\end{example}

We can now define TAS's.
\vspace{-1mm}
\begin{definition}
A \textbf{Tuple Artifact System} (TAS) is a triple $\Gamma = \langle \cala, \Sigma, \Pi \rangle$, 
where $\mathcal{A}$ is an artifact schema, $\Sigma$ is a set of services over $\mathcal{A}$, 
and $\Pi$, called the global pre-condition, is a condition over $\bar x$.
\end{definition}
\vspace{-1mm}

We next define the semantics of TAS. Intuitively, a run of a TAS on a database $D$ consists of 
an infinite sequence of transitions among artifact instances (also referred to as configurations, or snapshots), 
starting from an initial artifact tuple satisfying pre-condition $\Pi$. 
We begin by defining single transitions.

\vspace{-1mm}
\begin{definition}
Let $\Gamma = \langle \mathcal{A}, \Sigma, \Pi \rangle$ be a tuple artifact system, where
$\mathcal{A} = \langle \bar{x}, \mathcal{DB} \rangle$. We define the transition relation
among instances of $\cala$ as follows.
For two instances $(\nu, D), (\nu', D')$ and service $\sigma = \langle \pi, \psi, \bar y \rangle$,
$(\nu, D) \goto{\sigma} (\nu', D')$ if $D = D'$, $D \models \pi(\nu)$, $D \models \psi(\nu')$, and
$\nu'(y) = \nu(y)$ for each $y \in \bar{y}$.
\end{definition}
\vspace{-1mm}

Then a \emph{run} of the TAS $\Gamma = \langle \mathcal{A}, \Sigma, \Pi \rangle$ 
on database instance $D$ is an infinite sequence $\rho = \{ (I_i, \sigma_i) \}_{i \geq 0}$,
where each $I_i$ is an instance $(\nu_i, D)$ of $\cala$, $D \models \Pi(\nu_0)$, 
and for each $i > 0$, $I_{i-1} \goto{\sigma_{i}} I_{i}$.
In the run, $\sigma_0$ is a special initializing service {\em init}, whose role is to produce the instance $I_0$.

\vspace*{-1mm}
\subsection{Specifying Properties of TAS's with LTL-FO}
\vspace*{-1mm}

In this paper we focus on verifying temporal properties of runs of a tuple artifact system. 
For instance, in the business process of the example above, we would like to specify properties such as:
\vspace*{-1mm}
\begin{itemize} 
\item[$(\dag)$] If an order is taken and the ordered item is out of stock,
then the item must be restocked before it is shipped.
\end{itemize}
\vspace*{-1mm}

In order to specify such temporal properties we use, as in previous work, an extension of LTL (linear-time temporal logic).
LTL is propositional logic augmented with temporal operators such as {\bf G} (always),
{\bf F} (eventually), {\bf X} (next) and {\bf U} (until)
(e.g., see \cite{ltl}).  An LTL formula $\varphi$ with propositions $prop(\varphi)$ defines a property of sequences
of truth assignments to $prop(\varphi)$.  For example, ${\bf G}p$ says that $p$ always holds in the sequence, {\bf F}$p$
says that $p$ will eventually hold, $p {\bf U} q$ says that $p$ holds at least until $q$ holds, 
and ${\bf G}(p \rightarrow {\bf F}q)$ says that whenever $p$ holds, 
$q$ must hold later in the sequence. 

An LTL-FO property of a tuple artifact system $\cala$ is obtained 
starting from an LTL formula using some set $P \cup \Sigma$ of propositions.
Propositions in $P$ are interpreted as conditions over the variables $\bar{x}$ 
together with some additional \emph{global} variables $\bar y$,
shared by different conditions and allowing to refer to the state of the task at different moments in time.
The global variables are universally quantified over the entire property. 
A proposition $\sigma \in \Sigma$ indicates the application of service $\sigma$ in a given transition.  
LTL-FO formulas are defined as follows.
\vspace{-1mm}
\begin{definition}
Let $\Gamma = \langle \cala, \Sigma, \Pi \rangle$ be a TAS where $\cala = (\bar{x}, \db)$.
Let $\bar y$ be a finite sequence of variables in $\varid \cup \varnum$ disjoint from $\bar{x}$, called {\em global variables}.
An LTL-FO formula for $\Gamma$ is an expression $\forall \bar y \varphi_f$, where:
\vspace{-1mm}
\begin{itemize}\itemsep=0pt\parskip=0pt
\item $\varphi$ is an LTL formula with propositions $P \cup \Sigma$,
where $P$ is a finite set of proposition disjoint from $\Sigma$
\item $f$ is a function from $P$ to conditions over $\bar x \cup \bar y$
\item $\varphi_f$ is obtained by replacing each $p \in P$ with $f(p)$
\end{itemize}
\end{definition}
\vspace{-1mm}
For example, suppose we wish to specify property $(\dag)$.  The property is of the form
$\varphi = {\bf G} (p \rightarrow (\neg q \ {\bf U } \ r))$, which means:
if $p$ happens, then in the future $q$ will not happen until $r$ is true. Here $p$ says that 
the \textbf{EnterItem} service is called and chooses an out-of-stock item, $q$ states that 
the \textbf{ShipItem} service is called with the same item, and $r$ states that 
the service \textbf{Restock} is called to restock the item.
Since the item mentioned in $p$, $q$ and $r$ must be the same, 
the formula requires using a global variable $i$ denoting the ID of the item. 
This yields the following LTL-FO property: 
\vspace{-1mm}
\begin{align*}
\forall i ~~ \mathbf{G} (( {\mathtt{EnterItem}} \land \aitemid = i \land \ainstock = \text{``No''} ) \rightarrow \\
            ( \neg ({\mathtt{ShipItem}} \land \aitemid = i) {\ \bf U \ } ({\mathtt{Restock}} \land \aitemid = i)  ))
\end{align*}
A correct specification can enforce $(\dag)$ simply
by requiring in the pre-condition of \textbf{ShipItem} that the item is in stock.
One such pre-condition is $(\ainstock=\text{``Yes''} \land \astatus=\text{``Passed''})$,
meaning that the item is in stock and the customer passed the credit check.
However, in a similar specification where $\ainstock=\text{``Yes''}$ is not tested in the pre-condition
but performed in the post-condition of \textbf{ShipItem} 
(i.e. the post-condition requires that if $\ainstock=\text{``Yes''}$, then $\astatus$ stays unchanged
so the item is not shipped),
the LTL-FO property $(\dag)$ is violated because \textbf{ShipItem} can still be called without 
first calling the \textbf{Restock} service.
The verifier would detect this error and produce a counter-example illustrating the violation.




We say that a run $\rho = \{(I_i,\sigma_i)\}_{i \geq 0}$ satisfies 
$\forall \bar y \varphi_f$, where $prop(\varphi) = P \cup \Sigma$,
if $\varphi$ is satisfied, for all valuations of $\bar y$ in
$DOM_{id} \cup DOM_{val} \cup \{\anull\}$, by the sequence of truth assignments to $P \cup \Sigma$ induced by $f$ on the
sequence $\{(I_i,\sigma_i)\}_{i \geq 0}$. 
More precisely, for $p \in P$, the truth value induced for $p$ in $(I_i,\sigma_i)$ is the truth value of the condition $f(p)$
in $I_i$; a proposition $\sigma \in \Sigma$ holds in $(I_i,\sigma_i)$ if $\sigma_i = \sigma$.
A TAS $\Gamma$ satisfies $\forall \bar y \varphi_f(\bar y)$
if for every run $\rho$ of $\Gamma$ and valuation $\nu$ of $\bar y$, $\rho$ satisfies $\varphi_f(\nu(\bar y))$.

It is easily seen that for given $\Gamma$ with artifact variables $\bar x$ and LTL-FO formula  $\forall \bar y \varphi_f(\bar y)$, 
one can construct $\Gamma'$ with artifact variables $\bar x \cup \bar y$
such that $\Gamma \models \forall \bar y \varphi_f(\bar y)$ iff $\Gamma' \models \varphi_f$. 
Indeed, $\Gamma'$ simply adds $\bar y$ to the propagated variables in each service.
Therefore, we only consider in the rest of the paper quantifier-free LTL-FO formulas.

\vspace*{-2mm}
\section{The Spin-based Verifier} \label{sec:spin}

In this section we describe the implementation of SpinArt.
The implementation is based on Spin, the widely used model checker in software verification.
A brief review of Spin and \emph{Promela}, the specification language for Spin, is provided in Appendix \ref{app:promela}.

Building an artifact verifier based on Spin is a challenging task due to limitations of Spin and Promela.
In Promela, one can only specify variables with bounded domains ($\mathbf{byte}$, $\mathbf{int}$, etc.)
and bounded size (i.e. arrays with dynamic allocation are not allowed), 
but in the TAS model, the domains of the artifact variables and the database are unbounded
and the database instance can have arbitrary size, so a direct translation is not possible.
In addition, Spin cannot handle Promela programs of large size because 
the generated verifier $V$ would be too large for the C compiler.
Spin could also fail due to space explosion in the course of verification.
Thus, our implementation requires a set of nontrivial translations and optimizations, discussed next. 


\vspace*{-2mm}
\subsection{Symbolic Verification} \label{sec:verification}
The implementation makes use of the \emph{symbolic representation} technique developed in \cite{pods16}
to establish decidability and complexity results for HAS.
With the symbolic representation, the verification of TAS's is reduced to finite-state model checking 
that Spin can handle. 
Intuitively, given a TAS specification $\Gamma$ and an LTL-FO property $\varphi$,
we use {\em isomorphism types} to describe symbolically 
the structure of the portion of the database reachable from the current tuple of artifact variables by navigating the foreign keys.
An isomorphism type fully captures the information needed to evaluate any condition in $\Gamma$ and $\varphi$.
In addition, we can show, similarly to \cite{pods16}, that to check whether $\Gamma \models \varphi$,
it is sufficient to check that all \emph{symbolic runs} of isomorphism types satisfy $\varphi$, or equivalently, that no
symbolic run satisfies $\neg \varphi$. We define symbolic runs next.

We start by defining \emph{expressions}, which denote variables, constants and navigation via foreign keys 
starting from id variables. An expression is either:
\vspace{-1mm}
\begin{itemize}\itemsep=0pt\parskip=0pt
\item a constant $c$ in $\mathtt{const}(\Gamma,\varphi)$, the set of all constants that appear in $\Gamma$ or $\varphi$, or
\item a sequence $\xi_1.\xi_2.\ldots\xi_m$, where $\xi_1 = x$ for some id variable $x$, 
$\xi_2$ is an attribute of $R \in \db$ where $R.\emph{ID} = \type(x)$, 
and for each $i$, $2 \leq i < m$,
$\xi_i$ is a foreign key and $\xi_{i+1}$ is an attribute in the relation referenced by $\xi_i$.
\end{itemize}
\vspace{-1mm}

For a set of variables $\bar{y}$, we denote by $\cale(\bar{y})$ the set of expressions
$\{ y.w | y \in \bar y, |w| \geq 0\} \cup \mathtt{const}$. 
Such $\cale(\bar{y})$ for $\bar{y} \subseteq \bar{x}$ is called a {\em navigation set}.
Note that the length of expressions is bounded because of acyclicity of the foreign keys,
so $\cale(\bar{y})$ is finite.
We can now define isomorphism types.

\vspace{-1mm}
\begin{definition}
Let $\Gamma$ be a TAS with variables $\bar x$, and $\varphi$ an LTL-FO property of $\Gamma$.
An isomorphism type $\tau$ for $\Gamma, \varphi$, and variables $\bar y \subseteq \bar x$
consists of a navigation set $\cale(\bar y)$ together with an equivalence relation 
$\sim_\tau$ over $\cale(\bar y)$ such that:
\vspace{-1mm}
\begin{itemize}\itemsep=0pt\parskip=0pt
\item $c \not\sim_\tau c'$ for constants $c \neq c'$ in $\mathtt{const}(\Gamma,\varphi)$, and
\item if $u \sim_\tau v$ and $u.f, v.f \in \cale(\bar{y})$ then $u.f \sim_\tau v.f$.
\end{itemize}
\end{definition}
\vspace{-2mm}

We call an equivalence relation $\sim_\tau$ as above an {\em equality type} for $\tau$.
The relation $\sim_\tau$ is extended to tuples componentwise.
Intuitively, the second condition guarantees satisfaction of the key and foreign key dependencies.

\begin{figure}[!ht]
\centering
\includegraphics[width=0.35\textwidth]{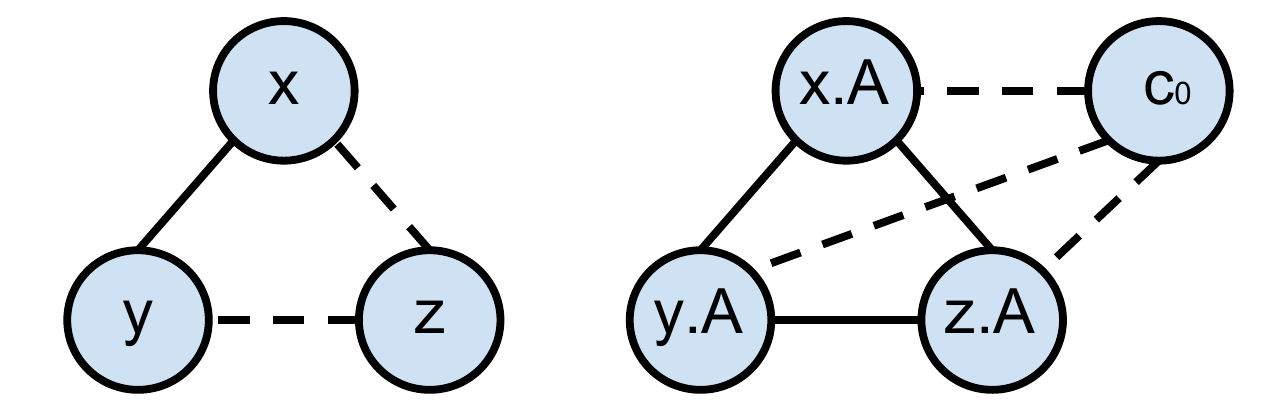}
\caption{An isomorphism type of variables $\{x, y, z\}$.}
\label{fig:isomorphism-type}
\end{figure}

\begin{example} \label{exm:partial}
Figure \ref{fig:isomorphism-type} shows an isomorphism types $\tau$ of variables $\{x, y, z\}$, where
$R(\emph{ID}, A)$ is the only database relation, $\{x, y, z\}$ are 3 variables of type 
$R.\emph{ID}$ and there is only one non-ID constant $c_0$. 
Each pair of expressions $(e, e')$ are connected with an solid line ($=$-edge) if $e \sim_\tau e'$
otherwise a dashed line ($\neq$-edge). 
The $\neq$-edges between $\{x, y, z\}$ and $\{x.A, y.A, z.A, c_0\}$ are omitted in the figure for clarity.
Note that since $(x, y)$ is connected with an $=$-edge,
$(x.A, y.A)$ must also be connected with $=$-edge as enforced by the key dependency.
\end{example}
\vspace*{-2mm}



Note that when $\bar y = \bar{x}$, $\tau$ provides enough information to evaluate conditions over $\bar{x}$.
Satisfaction of a condition $\varphi$ by an isomorphism type $\tau$, denoted $\tau \models \varphi$, is defined as follows:
\vspace{-1mm}
\begin{itemize}\itemsep=0pt\parskip=0pt
\item $x = y$ holds in $\tau$ iff $x \sim_\tau y$, 
\item $R(x, y_1, \dots, y_m)$ holds in $\tau$ for relation $R(\emph{ID}, A_1, \dots, A_m)$ iff 
$(y_1, \dots, y_m) \sim_\tau (x.A_1, \dots, x.A_m)$, and
\item Boolean combinations of conditions are standard.
\end{itemize}
\vspace{-1mm}

Let $\tau$ be an isomorphism type with navigation set $\cale(\bar{y})$ and equality type $\sim_\tau$.
The projection of $\tau$ onto a subset of variables $\bar{z}$ of $\bar{y}$, denoted as $\tau | \bar{z}$, 
is $(\sim_\tau | \bar{z}, \cale(\bar{z}))$ where $\sim_\tau|\bar{z}$ is the projection of $\sim_\tau$ onto $\cale(\bar{z})$. 
We define the symbolic transition relation among isomorphism types as follows: for a service $\sigma =  (\pi, \psi, \bar{y})$ in $\Sigma$,
$\tau \goto{\sigma} \tau'$ iff 
$\tau \models \pi$, $\tau' \models \psi$ and $\tau | \bar{y} = \tau' | \bar{y}$. 
\vspace{-1mm}
\begin{definition}
A symbolic run of $\Gamma = \langle \cala, \Sigma, \Pi \rangle$ 
is a sequence $\tilde{\rho} = \{(\tau_i, \sigma_i)\}_{i \geq 0}$ 
such that for each $i \geq 0$, $\tau_i$ is an isomorphism type, $\sigma_i \in \Sigma$, $\sigma_0 = \emph{init}$, 
$\tau_0 \models \Pi$ and $\tau_i \goto{~\sigma_{i+1}~} \tau_{i+1}$.
\end{definition}
\vspace{-2mm}

\begin{example} \label{exm:symbolic-transition}
Figure \ref{fig:symbolic-transition} shows an example of applying a symbolic transition on 
an isomorphism type. The previous isomorphism type $\tau$ (top-left) satisfies the pre-condition,
the next isomorphism type $\tau'$ (bottom) satisfies the post-condition, and they are
consistent in their projection to the propagated variables $\{x, z\}$ (top-right).
\begin{figure}[!ht]
\vspace{-6mm}
\centering
\includegraphics[width=1\textwidth]{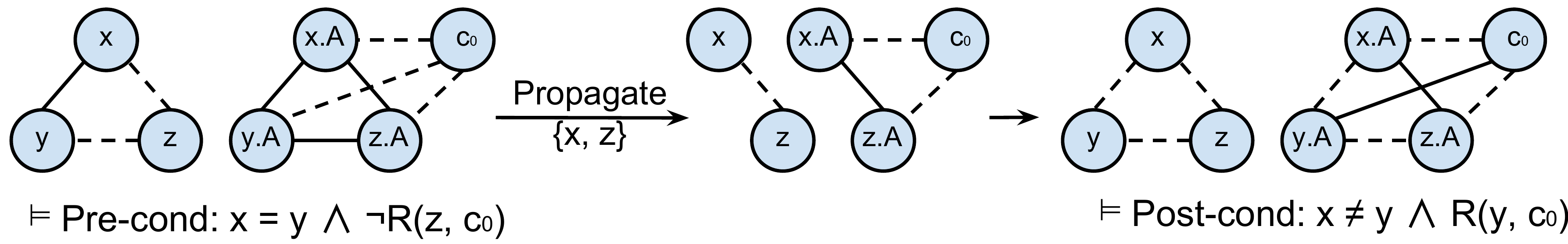}
\caption{Symbolic transition.}
\label{fig:symbolic-transition}
\end{figure}
\end{example}
\vspace{-1mm}

Satisfaction of a quantifier-free LTL-FO property on a symbolic run is defined in the standard way. 
One can show the following, similarly to \cite{pods16}.
\vspace*{-1mm}
\begin{theorem}\label{thm:symbolic}
Given a TAS $\Gamma$ and LTL-FO property $\varphi$ of $\Gamma$, $\Gamma \models \varphi$ iff 
for every symbolic run $\tilde{\rho}$ of $\Gamma$, $\tilde{\rho} \models \varphi$.
\end{theorem}
\vspace*{-1mm}


\vspace*{-2mm}
\subsection{Implementation of SpinArt} \label{sec:implementation}
Using Theorem \ref{thm:symbolic}, one can implement a verifier that constructs a Promela program 
$\calp$ to simulate the non-deterministic execution of symbolic transitions.
The program $\calp$ specifies $\cale(\bar{x})$ as its variables.
Each condition $\psi$ in $\Gamma$ and $\varphi$ is translated into a Promela condition $f(\psi)$ as follows.
\vspace{-1mm}
\begin{itemize}\itemsep=0pt\parskip=0pt
\item if $\psi = (x = y)$, then $f(\psi) = \psi$;
\item if $\psi= R(x, y_1, \dots, y_m)$ for relation $R(\emph{ID}, A_1, \dots, A_m)$, then
$f(\psi) = \bigwedge_{i = 1}^m (x.A_i = y_i)$;
\item Boolean connectives are handled in the standard way.
\end{itemize}
\vspace{-1mm}
\yuliang{removed one example}

Then $\calp$ simulates the following process of executing symbolic transitions.
First, $\calp$ initializes the constant expressions with distinct values and 
other expressions with non-deterministically chosen values that satisfy $f(\Pi)$. 
Then for each service $\sigma = (\pi, \psi, \bar{y})$, we construct a non-deterministic option with guard $f(\pi)$
that executes the following:
\vspace{-1mm}
\begin{itemize}\itemsep=0pt\parskip=0pt
\item[(i)] For each expression $e \in \cale(\bar{x}) - \cale(\bar{y})$, assign to $e$ a non-deterministically chosen value
from $\{0, \dots, |\cale(\bar{x})| - 1\}$.
\item[(ii)] Proceeds if $f(\psi)$ is $\mathtt{True}$ and for each pair of expressions $e$ and $e'$, $e = e'$ implies that
for every attribute $A$ where $\{e.A, e'.A\} \subseteq \cale(\bar{x})$, $e.A = e'.A$. Otherwise the run is blocked and invalidated.
\end{itemize}
\vspace{-1mm}
%
\begin{example}
First, each TAS condition is translated into a condition in Promela.
For example, the pre-condition in Example \ref{exm:symbolic-transition} is translated into
$$\texttt{ (x == y) \&\& !(z.A == c0)}.$$

Then, to construct the Promela program $\calp$, we first have a \textbf{do}-statement to 
ensure that the options constructed according to Sect. \ref{sec:implementation} are repeatedly chosen non-deterministically and executed. 
For example, the service in Example \ref{exm:symbolic-transition} is translated into the fragment of 
a Promela program shown in Fig. \ref{fig:translation}, 
where the \texttt{select(y : 0 .. N - 1)} statement is a built-in macro for 
assigning a variable with a value non-deterministically chosen
from a range (here $N$ is a constant equal to $|\cale(\bar{x})|$).

\begin{figure}[!ht]
\raggedleft
\begin{minipage}{1\textwidth}
\begin{lstlisting}
do
// check the pre-condition
:: ((x == y) && !(z.A == c0)) ->
   // choose values for y and y.A non-deterministically
   select(y : 0 .. N - 1); 
   select(y.A : 0 .. N - 1);
   // validate the post-condition
   if
   :: (x != y && y.A == c0) -> skip;
   fi;
   // validate the Keys and FKs
   if
   :: ((x != y || x.A == y.A) && (y != z || y.A == z.A) && (x != z || x.A == z.A)) -> skip;
   fi;
:: // another service
...
od
\end{lstlisting}
\end{minipage}
\caption{A fragment of a Promela program translated from a service.}
\label{fig:translation}
\end{figure}
\end{example}

Intuitively, each valid valuation $v$ to $\cale(\bar{x})$ corresponds to 
a valid isomorphism type $\tau$ of $\bar{x}$ where $e \sim_\tau e'$ iff $v(e) = v(e')$.
The guard ensures that the pre-condition holds.
Part (i) ensures that the set of next valuations covers all possible valid successors
of isomorphism types. Finally, the conditions in (ii) ensure that the post-condition holds
and the keys and FKs dependencies are satisfied in the next isomorphism type.

Finally, the LTL-FO formula $\varphi$ is translated into a LTL formula $\tilde{\varphi}$ in Promela by 
replacing each FO component $c$ with $f(c)$ defined above. 
The universally quantified variables of $\varphi$ are translated into extra variables added to the Promela program.
Small modifications to the LTL formula are also needed to skip 
the internal steps for assigning values and testing conditions in the run
such that the Spin verification only considers the snapshots right after complete service applications. We can show the following.
\vspace{-1mm}
\begin{lemma}\label{lem:promela}
Every symbolic run $\tilde{\rho} = \{(\tau_i, \sigma_i)\}_{i \geq 0}$ satisfies $\varphi$ iff $\calp \models \tilde{\varphi}$.
\end{lemma}
\vspace{-1mm}
The intuition of the above Lemma is that 
each valid valuation $v$ to $\cale(\bar{x})$ in $\calp$ corresponds to an unique isomorphism type $\tau$.
The translated transitions in Promela guarantees that the set of runs of $\calp$ captures the set of all symbolic runs.
So to check whether $\Gamma$ satisfies $\varphi$, 
it is sufficient to translate $(\Gamma, \varphi)$ into $(\calp, \tilde{\varphi})$ and 
verify whether $\calp \models \tilde{\varphi}$.

However, this approach is inefficient in practice for the following reasons. 
In part (ii), the size of the tests to ensure
satisfaction of the key and foreign key dependencies is quadratic in the number of expressions, 
so the compilation of $\calp$ and the generated verifier is slow or simply fails.
In (i), assigning to each $e$ values from $\{0, \dots, |\cale(\bar{x})| - 1\}$ is also infeasible 
because it leads to state explosion when the actual search is performed by the verifier.
As shown by the experiments, this leads to either slow execution or memory overflow. 
To overcome these two major obstacles, we introduce two key optimizations.
\vspace*{-2mm}
\subsection{Optimization with Lazy Dependency Tests} \label{sec:keys}
\vspace*{-1mm}
In the first optimization, we reduce the size of the generated Promela program by
eliminating the tests of key and foreign key dependencies in step (ii) of the above approach.
Instead, we introduce tests of the dependencies in a lazy manner, only when two expressions
are actually tested for equality. 
Formally, instead of performing the tests in (ii), we translate each condition $\psi$ of $(\Gamma, \varphi)$
into $f(\psi)$ then add the following additional tests: for every atom $(e = e')$ 
in the negation normal form\footnote{
With negations pushed down and merged with the $=$ and $\neq$ atoms, the only remaining Boolean operators are $\land$ and $\lor$.} of $f(\psi)$,
we replace $(e = e')$ with $\left( \bigwedge_{w : \{e.w, e'.w\} \subseteq \cale(\bar{x}) } e.w = e'.w \right)$
where $w$ is a sequence of attributes. 

The size of the tests in the resulting Promela program $\calp$ is 
$O((|\pi| + |\psi|) \cdot \max_{x \in \bar{x}} |\cale(x) | )$ for each service,
while the original size is $O(|\cale(\bar{x})|^2 \cdot a)$ 
where $a$ is the maximum arity in the database schema $\db$.
Typically, the size of a condition is much smaller than the number of expressions and 
$\max_{x \in \bar{x}} |\cale(x) |$ is also smaller than $|\cale(\bar{x})|$. 
We can see that the lazy dependency significantly reduces the size of the tests.

\vspace{-1mm}
\begin{example} \label{exm:ldt1}
Consider the database schema $\db = \{R(\emph{ID}, A, B), S(\emph{ID}, C, D)\}$ 
where $A$ and $B$ are foreign key attributes referencing the \emph{ID} of $S$
and $C, D$ are non-key attributes. A condition $R(x, y, z)$ is translated into 
\texttt{(x.A == y \&\& x.B == z)} without the optimization and 
\texttt{(x.A == y \&\& x.B == z \&\& x.A.C == y.C \&\& \\ x.A.D == y.D \&\& x.B.C == y.C \&\& x.B.D == y.D)}
if the lazy dependency tests optimization is applied. 
The additional terms in the conditions are added so that the tests for keys and FKs in the translation
can be removed.
\end{example}
\vspace{-1mm}

\begin{example} \label{exm:ldt2}
Consider the service and the translation shown in Fig. \ref{fig:translation}.
With lazy dependency tests, the translated pre-condition becomes
\texttt{(x == y) \&\& !(z.A == c0) \&\& (x.A == y.A)}
with one additional term \texttt{(x.A == y.A)}.
The translated post-condition is unchanged and the tests for keys and FKs
are removed (lines 12-14). The overall size of the translation is reduced.
\end{example}

\vspace*{1mm}
\noindent
\textbf{Correctness. } The modified translation using lazy dependency tests preserves correctness.
The intuition is the following. With the lazy tests, in some snapshot with valuation $v$ in the execution of $\calp$,
there could be two expressions $e, e'$ where $v(e) = v(e')$ and for some attribute $A$, $v(e.A) \neq v(e'.A)$,
but this does not matter because $e = e'$ is never tested during the current lifespan of $e$ and $e'$ 
(the segment of the symbolic run where $e$ and $e'$ are propagated), and neither are any of the prefixes of $e$ and $e'$.
So within the same lifespan, we are free to replace $v(e)$ and $v(e')$ 
with different values and the run of $\calp$ remains valid. 
Thus, there is no need to enforce the equality $e.A = e'.A$.

\vspace*{-2mm}
\subsection{Optimization with Assignment Set Minimization} \label{sec:spin-opt}
In the naive approach, assigning expressions with values chosen from a set of size $|\cale(\bar{x})|$
guarantees correctness by covering all possible isomorphism types, 
but it results in a large search space for Spin, which can lead to poor performance or memory overflow. 
The goal of this optimization is to reduce the size of the search space by 
minimizing the set of values used in the assignments while preserving the correctness of verification. 

We denote by $A(e)$ the \emph{assignment set} of a non-constant expression $e$, which is the set from which
the Promela program $\calp$ chooses non-deterministically values for $e$.
The technique relies on static analysis of $\calp$ and the translated property $\tilde{\varphi}$,
aiming to reduce the size of the assignment sets as much as possible.

The intuition behind the optimization is the following. 
We notice that searching for an accepting run in the generated Promela program $\calp$ can be regarded as
searching for a sequence of sets of constraints $\{C_i\}_{i \geq 0}$,
where each $C_i$ consists of the (in)equality constraints imposed on the current snapshot by the history of the run.
More precisely, the statements executed in $\calp$ can be divided into two classes:
(1) testing a condition $\pi$ and (2) assigning new values to some expressions.
At snapshot $i$, executing an (1)-statement can be viewed as adding $\pi$ to $C_i$
while $C_i$ should remain consistent (no contradiction implied by the $=$-or-$\neq$ constraints in $C_i$), and
a (2)-statement assigning a value to $e$ can be viewed as projecting away from $C_i$ constraints that involve $e$.
When we construct the assignment set $A(\cdot)$,
it is sufficient for correctness that the valuations generated with $A(\cdot)$ can witness the set of all reachable $C_i$'s,
which can be a small subset of all the possible isomorphism types. Thus, the resulting $A(\cdot)$ can be much smaller.

Computing all reachable $C_i$'s can be as hard as the verification problem itself. 
So instead, we over-approximate them with the \emph{constraint graph} $G$ of 
$(\calp, \tilde{\varphi})$ obtained by collecting all (in)equalities
from $(\calp, \tilde{\varphi})$, so that all $C_i$'s are subgraphs of $G$.

Formally, the constraint graph $G$ is an undirected labeled graph with $\cale(\bar{x})$ as the set of nodes, where
an edge $(e, e', \circ)$ is in $G$ for $\circ \in \{=, \neq\}$ if $(e \circ e')$ is an atom in any condition
of $\calp$ and $\tilde{\varphi}$ with all conditions converted in negation normal form. 

A subgraph $G'$ of $G$ is {\em consistent} if its edges do not lead to a contradiction (i.e., two nodes connected in $G'$ by a sequence of $=$-edges are not also connected by an $\neq$-edge). Observe that $G$ itself is generally not consistent, since it may contain
mutually exclusive constraints that never arise in the same configuration. On the other hand, each $C_i$ as above corresponds
to a consistent subgraph of $G$.

Intuitively, the approach to minimizing the assignment sets proceeds as follows. First,
consider the connected components of $G$ with respect to its equality edges. 
Clearly, distinct connected components can be consistently assigned disjoint sets of values.
Next, within each connected component, all expressions can be provided with the same assignment set, which we wish
to minimize subject to the requirement that it must provide sufficiently many values to satisfy each of its consistent subgraphs.

More precisely, we can show the following.
\vspace{-1mm}
\begin{lemma}\label{thm:chromatic}
Let $\calp'$ be the Promela program obtained from $(\calp, \tilde{\varphi})$ by replacing the assignment sets with any $A(\cdot)$ that satisfies:
\begin{enumerate}\itemsep=0pt\parskip=0pt
\item for every $(e, e', =) \in G$, $A(e) = A(e')$, and
\item for every consistent subgraph $G'$ of $G$, there exists a valuation $v$ such that for every $e \in \cale(\bar{x})$, $v(e) \in A(e)$ and for $\circ \in \{=, \neq \}$, $v(e) \circ v(e')$ if $(e, e', \circ) \in G'$.
\end{enumerate}
Then $\calp \models \tilde{\varphi}$ iff $\calp' \models \tilde{\varphi}$.
\end{lemma}
\vspace{-1mm}

Note that constants are not taken into account in the above lemma but can be included in a straightforward way.
Condition 2 implies that whenever a new valuation $v'$ is generated from a previous valuation $v$,
regardless of the previous and next constraint sets $C$ and $C'$, there exists a $v'$
that is consistent with $v$, $C$ and $C'$.

We next consider minimizing the assignment sets within each connected component.
It turns out that computing the minimal $A(\cdot)$ that satisfies the above conditions is closely related to computing 
the \emph{chromatic number} of a graph \cite{erdos1959graph}.
Recall that the chromatic number $\chi(G)$ of an undirected graph $G$ is the smallest number of colors needed to
color $G$ such that no two adjacent nodes share the same color. 
If the subgraph $G'$ in condition 2 is fixed, 
then the minimal $|A(\cdot)|$ is precisely the chromatic number of $G'$ restricted to only $\neq$-edges and 
with connected components of the $=$-edges merged into single nodes. We illustrate it with an example.

\begin{figure}[!t]
\centering
\includegraphics[width=0.8\textwidth]{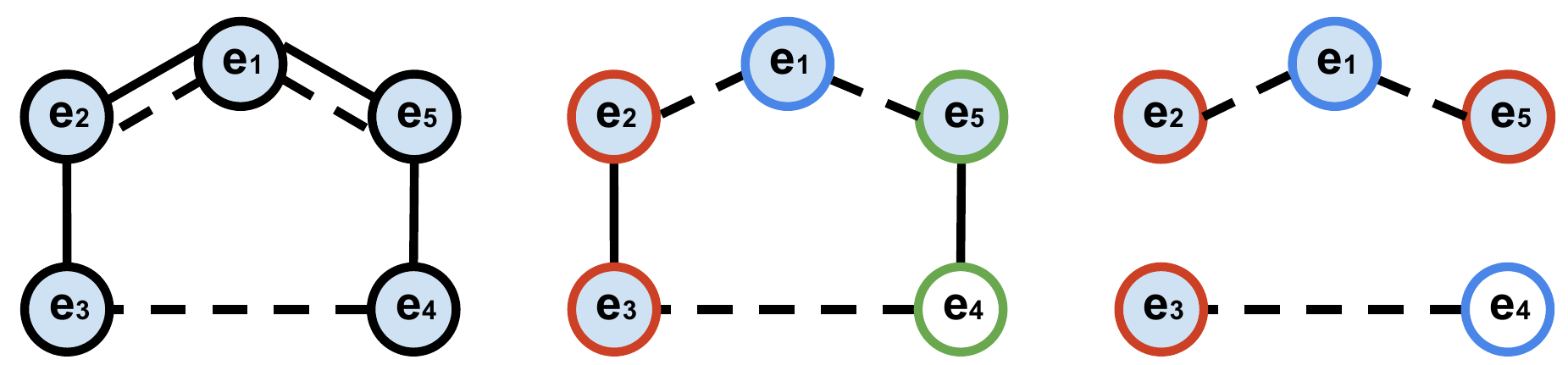}
\caption{Example of Assignment Sets Minimization.}
\label{fig:asm-example}
\end{figure}

\vspace{-1mm}
\begin{example}
Consider the constraint graph $G$ in the left of Fig. \ref{fig:asm-example}. 
The solid lines represent $=$-edges and the dashed lines represent $\neq$-edges.
The entire graph consists of a single
connected component of $=$-edges. 
To find the minimal $A(\cdot)$, we need to find the largest chromatic number over
all consistent subgraphs of $G$. Consider two consistent subgraphs $G_1$ (middle) and $G_2$ (right).
The chromatic number of $G_1$ is 3 because $(e_2, e_3)$ (and $(e_4, e_5)$) must share the same color, 
so $G_1$ is in fact a triangle.
The chromatic number of $G_2$ is 2 as it no long requires $e_2$ and $e_5$ to have different colors.
In fact, $G_1$ is the subgraph with the largest chromatic number, so setting $A(e_i) = \{0, 1, 2\}$ for every $i$
minimizes the assignment sets.
\end{example}
\vspace{-1mm}

As computing the chromatic number is {\sc np-hard}, it is not difficult to show that computing $A(\cdot)$ with minimal size
is also {\sc np-hard}. (We conjecture that it is $\mathrm{\Pi_2^P}$-{\sc hard}.) 
So computing the minimal $A(\cdot)$ can be inefficient. 
In the implementation, we use a simple algorithm that approximates the maximal chromatic number with the straightforward bound 
$\chi(G) (\chi(G) - 1) \leq 2m$ where $m$ is the number of $\neq$-edges within the connected component. 
The algorithm ensures satisfaction of the two conditions and produces reasonably small assignment sets in practice
because the constraint graph is likely to be very sparse and contains few $\neq$-edges. 
This is confirmed by our experiments.

\section{Experimental Results} \label{sec:experiment}

In this section we describe the experiments evaluating the performance of SpinArt.

\vspace*{1mm}
\noindent
\textbf{Benchmark. }
The benchmark used for the experiments consists of a collection of 
32 artifact systems modeling realistic business processes from different application domains.
Because of the difficulty in obtaining fully specified real-world data-driven business processes,
we constructed the benchmark starting from business processes specified
in the widely used BPMN model, that are provided by the official BPMN website \cite{bpmn}.
We rewrote the BPMN specifications into artifact systems
by manually adding the database schema, variables and service pre-and-post conditions.
Table \ref{tab:stat} provides some characteristics of the benchmark.

\vspace*{1mm}
\noindent
\textbf{LTL-FO Properties. }
On each workflow in the benchmark,
we run SpinArt on a collection of 12 LTL-FO properties constructed
using templates of real propositional LTL properties, yielding a total of 384 runs.
The LTL properties are all the 11 examples of safety, liveness and fairness properties collected from 
a standard reference paper \cite{sistla1994safety} and an additional property $\mathtt{False}$
used as a baseline when comparing the performance of SpinArt on different classes of LTL-FO properties.
We list all the templates of LTL properties in Table \ref{tab:ltlfo}.
We choose $\mathtt{False}$ as a baseline because it is the simplest property verifiable by Spin.  
By comparing the running time for a property with
the running time for $\mathtt{False}$ on the same specification,
we obtain the overhead for verifying the property.

\setlength{\tabcolsep}{3pt}
\begin{table}[!t]
\centering
\caption{Statistics of the BPMN benchmark.} \label{tab:stat}
\vspace*{-2mm}
  \begin{small}
  \begin{tabular}{ccccc}
    \toprule
\#Workflows   & Avg(\#Relations)  & Avg(\#Variables) & Avg(\#Services) \\
      \midrule
32     & 3.563      & 20.63   & 11.59 \\
    \bottomrule
  \end{tabular}
  \end{small}
\end{table}


For each workflow, we generate an LTL-FO property corresponding to each template by
replacing the propositions with FO conditions chosen from
the pre-and-post conditions of all the services and their sub-formulas.
Note that by doing so, the generated LTL-FO properties on the real workflows
are combinations of real propositional LTL properties and real FO conditions, and so
are close to real-world LTL-FO properties.

\vspace*{1mm}
\noindent
\textbf{Setup. } We implemented SpinArt in \texttt{C++} with Spin version 6.4.6.
All experiments were performed on a Linux server with a quad-core Intel i7-2600 CPU and 16G memory.
To allow larger search space, Spin was run with the state compression optimization turned on.
For faster execution, the Spin-generated verifier was compiled with $\mathtt{gcc}$ and the -O2 optimization.
The time and memory limit of each run was set to 10 minutes and 8G respectively.

\vspace*{1mm}
\noindent
\textbf{Performance. }
In addition to running the full verifier (SpinArt-Full), we also ran
the verifier with the lazy dependency tests optimization (LDT) turned off (SpinArt-NoLDT) and
with assignment set minimization (ASM) turned off (SpinArt-NoASM).
For all the verifiers, we compare their number of failed runs (timeout or memory overflow),
the average compilation time\footnote{All averages (running times and \#States) are taken over the successful runs.}
for generating the executable verifier (Compile-Time),
the average execution time of the generated verifier (Verify-Time),
the average total running time (Verify-Time + Compile-Time),
and the average number of reached states as reported by Spin.

\setlength{\tabcolsep}{4pt}
\begin{table}[!t]
  \centering
  \caption{Performance of SpinArt in different modes.}
  \vspace*{-3mm}
  \label{tab:performance}
  \begin{small}
  \begin{tabular}{ccccccc}
    \toprule
    Mode     & {\#Failed-Runs} & Total-Time & Verify-Time & Compile-Time & \#States \\ \midrule
    SpinArt-NoASM & 48 / 384    & 21.399s     & 14.379s  & 7.020s  & 1,547,211 \\
    SpinArt-NoLDT & 3 / 384     & 12.240s     & 3.769s  & 8.471s & 809,025 \\
    SpinArt-Full & \textbf{3} / 384     & \textbf{2.970s} & \textbf{0.292s} & \textbf{2.678s} & \textbf{44,826} \\
    \bottomrule
  \end{tabular}
  \end{small}
\vspace*{-5mm}
\end{table}

The results are shown in Table \ref{tab:performance}.
We can see that the performance of SpinArt is promising.
Its average total running time is within 3 seconds and
there are only 3/384 failed runs ($<$1\%) due to memory overflow.
This is a strong indication that the approach is sufficiently practical for real-world workloads.
The full verifier is also significantly improved compared to SpinArt-NoLDT and SpinArt-NoASM.
Without ASM, the the verifier failed on 12.5\% (48/384) of all runs and
the average running time is $>$7x times faster when the optimization is turned on.
Without LDT, most of the runs are still successful,
but the average total running time is $>$4 times faster with the optimization turned on.
Both optimizations significantly reduce the size of the state space ($>$95\% in total),
resulting in much shorter verification time.

We next discuss the effect of each optimization in more detail.

\begin{figure}[t!]
\begin{minipage}[b]{.48\textwidth}
\centering
\includegraphics[width=1\textwidth]{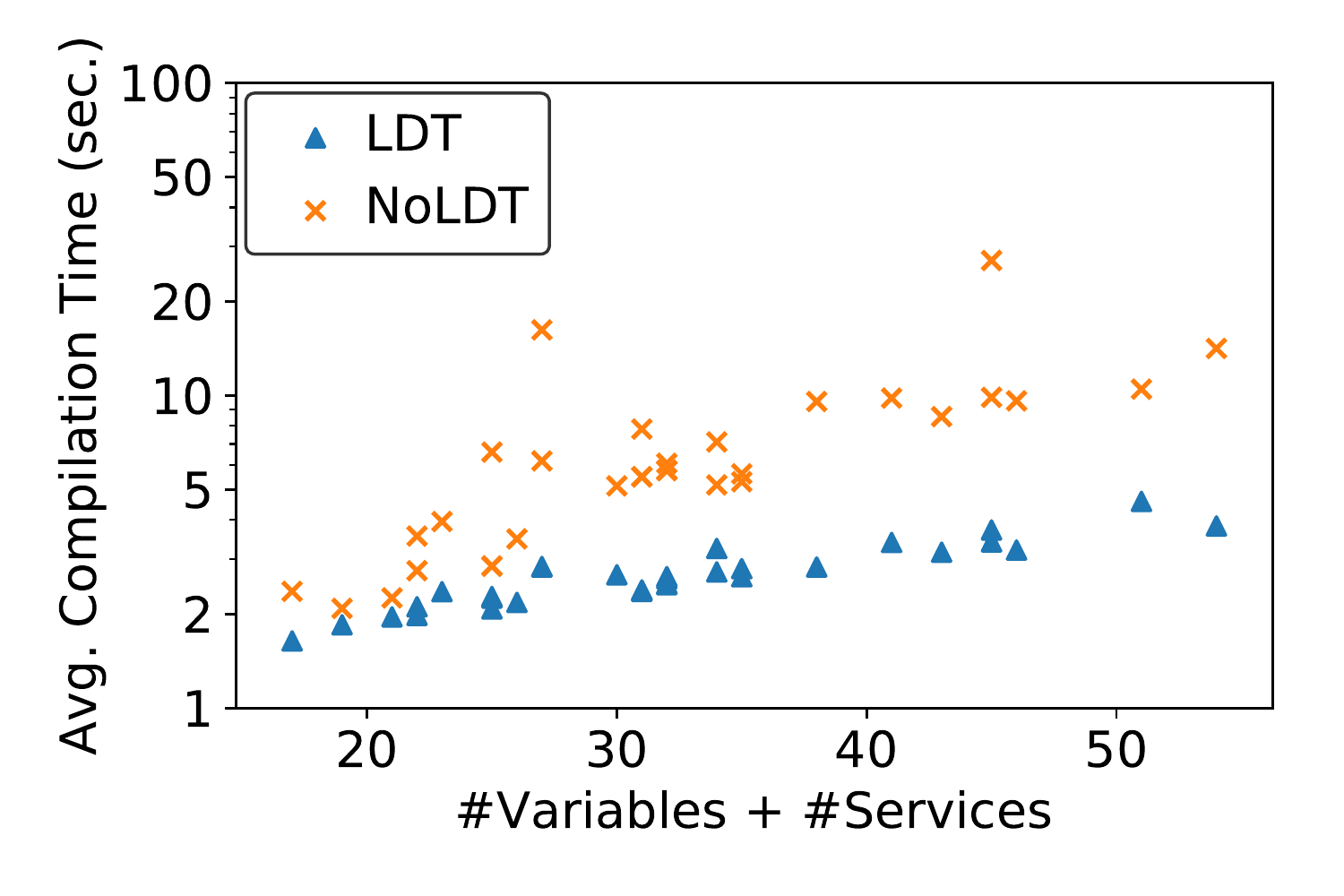}
\vspace*{-5mm}
\caption{Compilation time with or without Lazy Dependency Tests.\label{fig:ldt}}
\end{minipage}
\hfill
\begin{minipage}[b]{.48\textwidth}
\centering
\includegraphics[width=1\textwidth]{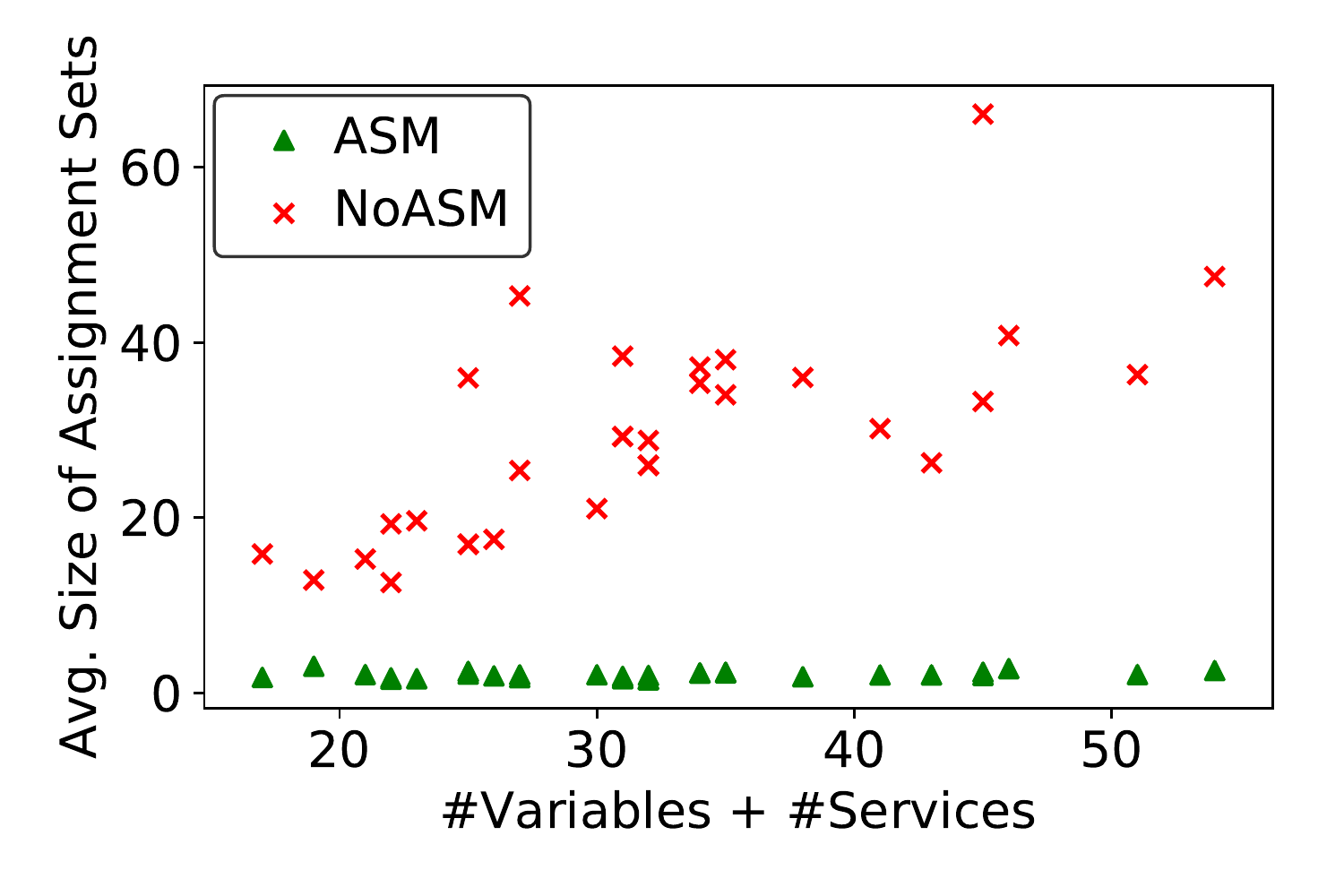}
\vspace*{-5mm}
\caption{Average size of the assignment sets with or without minimization.\label{fig:asm}}
\end{minipage}
\end{figure}

\vspace*{1mm}
\noindent
\textbf{Effect of Lazy Dependency Tests. }
From Table \ref{tab:performance}, we observe that for the successful runs,
compilation time accounts for a large fraction of the total running time,
so minimizing the size of the Promela program is critical to improve the overall performance of a Spin-based verifier.
Figure \ref{fig:ldt} shows the changes in the compilation time as
the size of the input specification (\#Variables + \#Services) increases,
for runs with or without the LDT optimization. Each point in the figure
corresponds to one specification and the compilation time is measured by the average
compilation time of all runs of the specification.
The figure shows that with LDT, the compilation time grows not only slower as the input size increases,
but in some cases it can compile $>$10 times faster than compilation without LDT.
Overall, LDT leads to an average speedup of {3.2x} in compilation.


\vspace*{1mm}
\noindent
\textbf{Effect of Assignment Set Minimization. }
We show the effectiveness of Assignment Set Minimization (ASM) by
comparing the approximation algorithm for ASM with a na\"{i}ve approach (NoASM) where the size of the assignment set of each expression $e$
is simply set to the number of expressions having the same type as $e$.
Figure \ref{fig:asm} shows the growth of the average size of the assignment sets as
the size of the input specification increases.
For ASM, the average size stays very low ({2.05} in average) as the input size
grows. This shows that our algorithm is near-optimal in practice.
Compared to the naive approach where the average size increases linearly with the input size,
our approach produces much smaller assignment sets.
In some cases, the assignment set generated by the algorithm is $>$30 times smaller than
the ones generated by the naive approach.

\setlength{\tabcolsep}{2pt}
\begin{table}[!ht]
\centering
\caption{Average running time of verifying different classes of LTL-FO properties.}
\vspace*{-2mm}
\label{tab:ltlfo}
  \begin{small}
  \begin{tabular}{ccc|ccc}
    \toprule
      Templates  & Avg(Time) & Overhead & Templates  & Avg(Time) & Overhead \\
      \midrule
$\mathtt{False}$  & 2.68s & 0.00\% & $\mathbf{G} ( \varphi \rightarrow \mathbf{F} \psi )$  & 2.72s & 1.45\% \\
$\mathbf{G} \varphi$  & 2.68s & -0.26\% & $\mathbf{F} \varphi $  & 2.80s & 4.08\% \\
$(\neg \varphi \ \mathbf{U} \ \psi )$  & 2.70s & 0.61\% & $\mathbf{GF} \varphi \rightarrow \mathbf{GF} \psi$  & 2.91s & 9.36\% \\
$(\neg \varphi \mathbf{U} \psi ) \land \mathbf{G} (\varphi \rightarrow \mathbf{X}(\neg \varphi \mathbf{U} \psi ) )$  & 5.07s & 70.02\% & $\mathbf{GF} \varphi $  & 3.07s & 15.14\% \\
$\mathbf{G} (\varphi \rightarrow ( \psi \lor \mathbf{X} \psi \lor \mathbf{X} \mathbf{X} \psi) )$  & 2.72s & 1.40\% & $\mathbf{G} (\varphi \lor \mathbf{G} \psi)$  & 2.71s & 0.85\% \\
$\mathbf{G}(\varphi \lor \mathbf{G}(\neg \varphi))$  & 2.69s & 0.28\% & $\mathbf{FG} \varphi \rightarrow \mathbf{GF} \psi$  & 2.91s & 9.11\% \\
\bottomrule
  \end{tabular}
  \end{small}
\vspace*{-5mm}
\end{table}

\vspace*{1mm}
\noindent
\textbf{Effect of the Structure of LTL-FO Properties. }
Next, we measure the performance on different classes of LTL-FO properties.
Table \ref{tab:ltlfo} lists all the LTL templates used in generating the LTL-FO properties
and their intuitive meaning, as in \cite{sistla1994safety}.
For each template, we measure the average running time over all runs with LTL-FO properties
generated using the template. In addition, we measure the overhead of verifying
a LTL-FO property by comparing with its running time for the property $\mathtt{False}$,
the simplest non-trivial property for SpinArt.
The overhead of a class of LTL-FO properties is obtained by the average overhead of
all properties of the same class. The result in Table \ref{tab:ltlfo} shows that
the average running time stays within 2x of the average running time for $\mathtt{False}$ and
the maximum average overhead is about 70\%.
The overhead increases as the LTL property becomes more complex, but is within a reasonable range.
Note that this is much better than the theoretical upper bound,
which is exponential in the size of the LTL formula.

\yuliang{I think the following can be moved to the appendix.}

\vspace*{1mm}
\noindent
\textbf{Results on Synthetic Workflows. }
Finally, we stress-test the performance of SpinArt by running it on a set of 120 randomly generated TAS specifications.
All components of each specification were generated fully at random for a specified size. 
Each specification has 5 relations in the DB schema, 75 variables and 75 services with randomly generated pre-and-post conditions.
The ones with empty search space due to unsatisfiable conditions were removed from this benchmark.
On each workflow, we ran SpinArt to verify 12 LTL-FO properties generated from the templates in Table \ref{tab:ltlfo}, resulting in
1440 runs in total. Among these runs, SpinArt succeeded in 1000/1440 ($\sim$70\%) runs with an average running time of \textbf{83.983s}.
The remaining runs failed due to timeout or memory overflow. 
As preformance remains acceptable on the much larger synthetic workflows, 
the results suggest that SpinArt is scalable to complex workflows.
Note that the two optimizations are essential to the above results, 
since almost all runs failed due to compiler crash if either optimization is turned off.

\section{Additional Related Work}\label{sec:related}


The artifact verification problem has been studied mainly from a theoretical perspective.
As mentioned in Sect. \ref{sec:intro}, fully automatic artifact verification is a challenging problem
due to the presence of unbounded data. To deal with the resulting infinite-state system, 
a symbolic approach was developed in \cite{DHPV:ICDT:09} allowing a reduction to finite-state model checking and
yielding a {\sc pspace} verification algorithm for the simplest variant of the
model (no database dependencies and uninterpreted data domain).  
\cite{tods12} extended this approach to allow for database dependencies and 
numeric data testable by arithmetic constraints. 
The symbolic approach developed in \cite{DHPV:ICDT:09,tods12} and revisited in HAS \cite{pods16} 
provides the theoretical foundation of our Spin-based implementation.

Another line of work considers the verification problem for runs starting from a {\em fixed} initial database.
During the run, the database may evolve via updates, insertions and deletions. 
Since inputs may contain fresh values from an infinite domain, 
this verification variant remains infinite-state.
The property languages are fragments of first-order-extended $\mu$-calculus~\cite{DBLP:journals/ijcis/GiacomoMR12}.
Decidability results are based on sufficient syntactic restrictions 
~\cite{DBLP:journals/ijcis/GiacomoMR12,DBLP:conf/pods/HaririCGDM13,calvanese2015verification}. 
\cite{DBLP:conf/icsoc/BelardinelliLP12} derives decidability of the verification variant
by also disallowing unbounded accumulation of input values, 
but this condition is postulated as a semantic property 
(shown undecidable in \cite{DBLP:conf/pods/HaririCGDM13}). 
\cite{abdulla2016recency} takes a different approach, in which
decidability is obtained for {\em recency-bounded} artifacts, 
in which only recently introduced values are retained in the current data.

On the practical side of artifact verification, 
\cite{rawsys} specifies business processes in 
a Petri-net-based model extended with data and process components, 
in the spirit of the theoretical work of 
\cite{rosa2011decidability,badouel2015petri,lazic2008nets,sidorova2011soundness}, 
which extends Petri nets with data-carrying tokens.
The verifier of \cite{rawsys} differs fundamentally from ours in that 
properties are checked only for a given initial database, whereas 
our verifier checks properties {\em regardless} of the initial database.
\cite{gsmc,gonzalez2012verifying,gonzalez2013model} implemented 
a verifier for artifact systems specified directly in the GSM model.
While the above models are expressive, the verifiers require
restrictions strongly limiting modeling power \cite{gonzalez2012verifying}, 
or predicate abstraction resulting in loss of soundness and/or completeness \cite{gsmc,gonzalez2013model}. 
Lastly, the properties verified in \cite{gsmc,gonzalez2013model} focus on 
temporal-epistemic properties in a multi-agent finite-state system. 
Thus, the verifiers in these works have a different focus and are incomparable to ours.
Practical verification has also been studied in business process management
(see \cite{bpm-verification} for a survey). The considered models are mostly process-driven 
(BPMN, Workflow-Net, UML etc.), with the business-relevant data abstracted away.
The implementation of a verifier for data-driven web applications was studied in \cite{wave2005} and \cite{wave2006}.
The model is similar in flavor to the artifact system model but incomparable due to the different application domains.
An attempt to build a verifier based on Spin was made in \cite{wave2005} but failed due to search space explosion,
confirming that the optimizations used in our implementation of SpinArt are essential.

\vspace*{-2mm}
\section{Conclusion, Related Work and Discussion} \label{sec:conclusion}
\vspace*{-2mm}

%
We reported on our implementation of SpinArt, a verifier for data-driven workflows using 
the widely used off-the-shelf model checker Spin.
With a translation based on the symbolic representation developed in \cite{pods16} enhanced with nontrivial optimizations,
SpinArt achieves good performance on a realistic business process benchmark.
We believe this is a first successful attempt to bridge the gap between theory and practice 
in verification of data-driven workflows, with full support for unbounded data and relying on an off-the-shelf model checker.

\yuliang{The following paragraph can be shortened.}

\vspace*{1mm}\noindent
\textbf{Discussion.} The focus of our work is on sound and complete 
artifact verifiers, in contrast to incomplete verifiers (e.g. based on theorem provers).
Within this scope, 
SpinArt establishes a practical trade-off point on the
spectrum ranging from using off-the-shelf general software
verifiers to developing dedicated verifiers from scratch.

On the one hand, off-the-shelf tools share a number of limitations
which are inherited by verifiers based on them (including ours).
For instance, general-purpose model checkers have limited support for unbounded data. 
While our work mitigates this limitation by supporting 
the unbounded read-only database with symbolic representation, our model does not support
other ingredients of the HAS (and GSM) model, such as 
dynamically updatable artifact relations, because
they require an enhanced symbolic representation counting 
the number of tuples of different isomorphism types,
which exceeds the capabilities of Promela/Spin.

On the other hand, from-scratch implementation is costly
as it duplicates functionality already present in mature tools such as Spin.
More importantly, the initial implementation cost is typically outweighed by maintenance cost over
the verifier's lifetime. In contrast, verifiers based on off-the-shelf model checkers
feature lower development and maintenance cost. 



%

\bibliographystyle{splncs}
\bibliography{reference,art2,artifacts}

\appendix

\section{Review of Spin and Promela} \label{app:promela}
\lstset{language=Promela}

The implementation of our artifact verifier relies on Spin, a widely used model checker in software verification.
Spin supports the verification of LTL properties of models specified in \emph{Promela}, 
a C-like modeling language for parallel systems. 
At a high level, a single-process Promela program can be viewed as a non-deterministic C program,
where one can specify variables of fixed bit-length (e.g. $\mathbf{byte}$, $\mathbf{short}$, $\mathbf{int}$)
and statements that manipulate the variables (e.g. assignments, goto, etc.).
Non-determinism is specified using the \textbf{if}- and \textbf{do}-statements illustrated in Fig. \ref{fig:promela}.

\begin{figure}
\vspace*{-6mm}
\begin{minipage}{.5\textwidth}
\begin{lstlisting}
if
:: (a == 0) -> b = a + 1;
:: (b > 1) -> c = a; 
:: a = a - 1; 
   b = b + 1;
fi
\end{lstlisting}
\end{minipage}\hfill
\begin{minipage}{.5\textwidth}
\begin{lstlisting}
do
:: count = count - 1;
:: a = a + 2;
:: (count == 0) -> break;
:: (count > 0) -> skip;
od
\end{lstlisting}
\end{minipage}
\caption{Examples of Promela program (Left: \textbf{if}-statement; Right: \textbf{do}-statement).}\label{fig:promela}
\end{figure}

When the \textbf{if}-statement is executed, one of its options with no guard or
with its guard evaluating to $\mathtt{True}$ is chosen non-deterministically and executed.
Each option is a sequence of one or more statements.
If no option can be chosen, then the run blocks the is not considered as a valid run when Spin is executed.
The \textbf{do}-statement is similar to the \textbf{if}-statement, with the difference that 
the execution is repeated after an option is completed.
Nesting is allowed within the \textbf{if}- or \textbf{do}-statements.

Developers can verify LTL properties of a Promela program using Spin.
Given a Promela program $\calp$, a developer can write LTL properties 
where the propositions are Boolean conditions over the variables of $\calp$,
such as: 
``\texttt{G ((a == 1) -> F (b > 0 || c < 0))}''.

To check satisfaction of a LTL property $\varphi$, 
Spin first produces the source code of a problem-specific verifier $V$ in C.
Then $V$ is compiled with a C-compiler (e.g. $\mathtt{gcc}$) and executed to produce the result.

\end{document}